# Improved Adaptive Sparse Channel Estimation Using Re-Weighted L1-norm Normalized Least Mean Fourth Algorithm

Chen Ye[†1], Guan Gui[1], Li Xu[1] and Nobuhiro Shimoi[2]

1. Department of Electronics and Information Systems, Akita Prefectural University, Yurihoujo, 015-0055 Japan
2. Department of Machine Intelligence and Systems Engineering, Akita Prefectural University, Yurihoujo, 015-0055 Japan

(Tel: +81-184-27-2241; E-mail: guiguan@akita-pu.ac.jp)

**Abstract:** In next-generation wireless communications systems, accurate sparse channel estimation (SCE) is required for coherent detection. This paper studies SCE in terms of adaptive filtering theory, which is often termed as adaptive channel estimation (ACE). Theoretically, estimation accuracy could be improved by either exploiting sparsity or adopting suitable error criterion. It motivates us to develop effective adaptive sparse channel estimation (ASCE) methods to improve estimation performance. In our previous research, two ASCE methods have been proposed by combining forth-order error criterion based normalized least mean fourth (NLMF) and L1-norm penalized functions, i.e., zero-attracting NLMF (ZA-NLMF) algorithm and reweighted ZA-NLMF (RZA-NLMF) algorithm. Motivated by compressive sensing theory, an improved ASCE method is proposed by using reweighted L1-norm NLMF (RL1-NLMF) algorithm where RL1 can exploit more sparsity information than ZA and RZA. Specifically, we construct the cost function of RL1-NLMF and hereafter derive its update equation. In addition, intuitive figure is also given to verify that RL1 is more efficient than conventional two sparsity constraints. Finally, simulation results are provided to correlate this study.

## 1. INTRODUCTION

Broadband signal transmission is becoming one of the mainstream techniques in the next generation communication systems [1]–[3]. Due to the fact that frequency-selective channel fading is unavoidable, accurate channel state information (CSI) is necessary at the receiver for coherent detection [4]. One of effective approaches is adopting adaptive channel estimation (ACE). A typical framework of ACE is shown in Fig. 1. It is well known that ACE using least mean fourth (LMF) algorithm outperforms the least mean square (LMS) algorithm in achieving a good balance between convergence and steady-state performances [5]. However, standard LMF algorithm is unstable for that its stability depends on the following three factors: input signal power, noise power and weight initialization [5]. To improve the stability of LMF, stable normalized LMF (NLMF) algorithm was proposed in [6][7].

However, standard NLMF algorithm based ACE does not consider channel structure which could be utilized to improve estimation accuracy. Recently, many channel measurement experiments have verified that broadband channels often exhibit sparse structure as shown in Fig. 2. In other words, sparse channel is consisted of a very few channel coefficients and most of them are zeros [8]–[10].

To estimate the sparse channel, two ASCE methods were proposed by incorporating sparse constraint into NLMF algorithm, i.e., zero-attracting NLMF (ZA-NLMF) and reweighted ZA-NLMF (RZA-NLMF) [11]. According to compressive sensing (CS) [12], an improved ASCE method using re-weighted $L_1$-norm NLMF (RL1-NLMF) algorithm is proposed. The contribution of this paper is briefly summarized as follows. Firstly cost function of RL1-NLMF is constructed and then update equation is derived. Secondly to evaluate the sparse constraint strength of the RL1, intuitive figure is depicted to compare with ZA and RZA. By virtue of Monte Carlo (MC) measurement approach, at last mean square deviation (MSD) performance curves are depicted to verify the effectiveness of proposed method in scenarios therein different step-sizes, different channel sparsity as well as different SNR regimes.

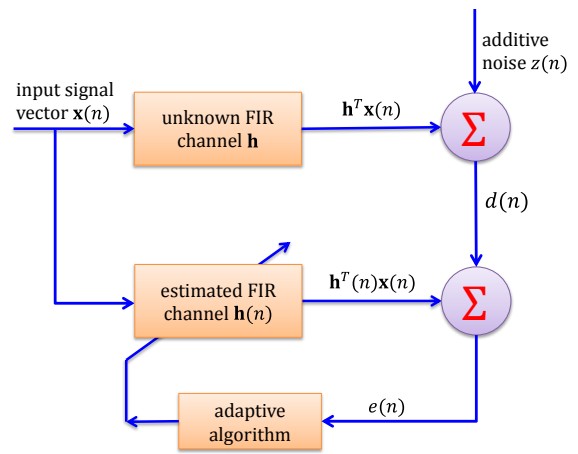

Fig. 1. ASCE for broadband communication systems.

† Chen Ye is the presenter of this paper.

This paper is organized as follows. A system model is described and standard LMF and NLMF algorithms are introduced in Section 2. In section 3, ZA-NLMF and RZA-NLMF are reviewed and improved ASCE using RL1-NLMF is proposed. Simulation results are presented in Section 4 in order to evaluate the proposed method. Finally, we conclude this paper in Section 5.

*Notation*: Throughout the paper, capital bold letters and small bold letters denote matrices and row/column vectors, respectively; The superscripts $(\cdot)^T$ denotes the transpose; $\mathrm{E}(\cdot)$ denotes the expectation operator; $\|\boldsymbol{h}\|_p$ stands for the Lp-norm operator which is computed as $\|\boldsymbol{h}\|_p = (\sum_i |h_i|^p)^{1/p}$, where $p \in \{1, 2\}$.

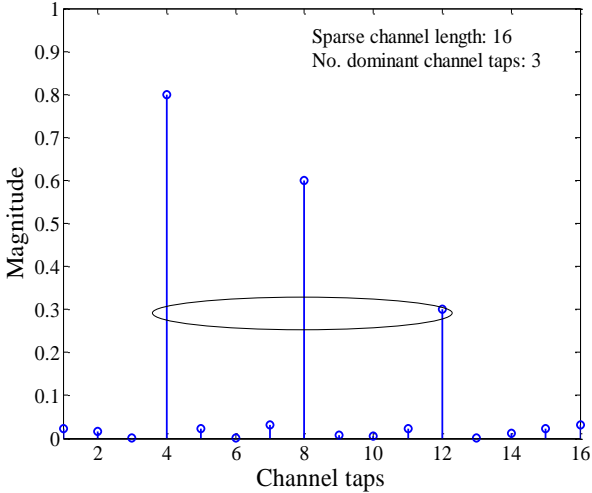

Fig. 2 A typical example of sparse multipath channel.

## 2. SYSTEM MODEL AND STANDARD NLMF ALGORITHM

Consider a baseband frequency-selective fading wireless communication system where finite impulsive response (FIR) of sparse channel vector $\boldsymbol{h} = [h_1, h_2, ..., h_{FIR}]^T$ which is supported only by $K$ nonzero channel dominant taps. Assume that an input training signal $\boldsymbol{x}(n)$ is used to probe the unknown sparse channel. At the receiver side, observed signal $d(n)$ is given by

$$d(n) = \boldsymbol{h}^T \boldsymbol{x}(n) + z(n), \quad (1)$$

where $\boldsymbol{x}(n) = [\boldsymbol{x}_1(n), \boldsymbol{x}_2(n), ..., \boldsymbol{x}_{FIR}(n)]^T$ denotes the training signal vector, and $z(n)$ is the additive white Gaussian noise (AWGN) which is assumed to be independent with $\boldsymbol{x}(n)$. The objective of ASCE is to adaptively estimate the unknown sparse channel estimator $\boldsymbol{h}(n)$ using the training signal $\boldsymbol{x}(n)$ and the observed signal $d(n)$. According to [5], standard LMF algorithm based adaptive channel estimation, where the cost function can be constructed as

$$G_{LMF}(n) = \frac{1}{4} e^4(n), \quad (2)$$

where $e(n) = d(n) - \boldsymbol{h}^T(n)\boldsymbol{x}(n)$ is $n$-th adaptive updating error. Based on Eq. (2), the LMF based adaptive filtering algorithm can be derived as

$$\begin{aligned} \boldsymbol{h}(n+1) &= \boldsymbol{h}(n) + \mu \frac{\partial G_{LMF}(n)}{\partial \boldsymbol{h}(n)} \\ &= \boldsymbol{h}(n) + \mu e^3(n)\boldsymbol{x}(n), \end{aligned} \quad (2)$$

where $\mu$ denotes the step-size of gradient. Since LMF algorithm is unstable in adaptive updating process and hence it hard to be employed in channel estimation directly [2]. To improve the algorithm's reliability, normalized LMF (NLMF) algorithm was proposed in [7]. The updating equation is given by

$$\begin{aligned} \boldsymbol{h}(n+1) &= \boldsymbol{h}(n) + \mu \frac{e^3(n)\boldsymbol{x}(n)}{\|\boldsymbol{x}(n)\|_2^2 \left(\|\boldsymbol{x}(n)\|_2^2 + e^2(n)\right)} \\ &= \boldsymbol{h}(n) + \mu_N \frac{e(n)\boldsymbol{x}(n)}{\|\boldsymbol{x}(n)\|_2^2}, \end{aligned} \quad (3)$$

where

$$\mu_N = \frac{\mu e^2(n)}{\|\boldsymbol{x}(n)\|_2^2 + e^2(n)}, \quad (4)$$

denotes variable step-size which depends on initial step-size $\mu$, updating error $e(n)$ and input signal $\boldsymbol{x}(n)$. Based on the standard NLMF algorithm (4), we will review two ASCE methods using ZA-NLMF and RZA-NLMF and proposed an improved ASCE using RL1-NLMF algorithm.

## 3. ASCE METHODS

### 3.1. ASCE using ZA-NLMF

Recall that the adaptive channel estimation method uses standard NLMF algorithm in Eq. (4), however, the standard linear method does not take advantage of the channel sparsity. It was caused by its original cost function in (2) which does not utilize the sparse constraint or penalty function. Hence here L1-norm sparse constraint to the cost function in (4) is introduced then obtain a new cost function as follow

$$G_{ZA}(n) = \frac{1}{4} e^4(n) + \lambda_{ZA} \|\boldsymbol{h}(n)\|_1, \quad (5)$$

where $\lambda_{ZA}$ denotes a regularization parameter which balances the mean-fourth error term and sparsity of $\boldsymbol{h}$. The updated equation of ZA-NLMF algorithm [13] is given as follow

$$\boldsymbol{h}(n+1) = \boldsymbol{h}(n) + \mu_N \frac{e(n)\boldsymbol{x}(n)}{\|\boldsymbol{x}(n)\|_2^2} - \gamma \operatorname{sgn}(\boldsymbol{h}(n)), \quad (6)$$

where $\gamma = \mu\lambda_{ZA}$ and $\operatorname{sgn}(\cdot)$ denotes the sign function which is defined as follows

$$\operatorname{sgn}(h_i(n)) = \frac{\partial \|h_i(n)\|_1}{\partial h_i(n)} = \begin{cases} 1, & h_i(n) > 0 \\ 0, & h_i(n) = 0, \\ -1, & h_i(n) < 0 \end{cases} \quad (7)$$

for $i \in \{1,2,...,FIR\}$. It is well known that ZA-NLMF can be applied in sparse channel estimation but the sparsity penalty is inefficient [14].

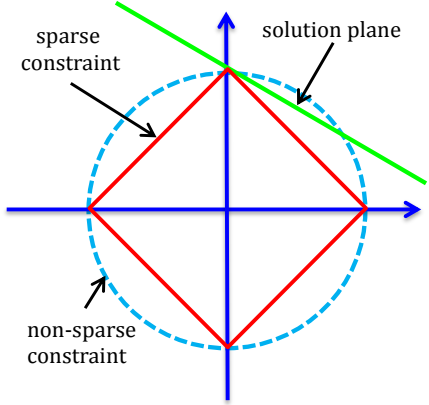

Fig. 4. Difference between L2-norm solution and L1-norm one.

### 3.2. ASCE using RZA-NLMF

Motivated by reweighted L1-minimization sparse recovery algorithm [6] in CS [15], an improved ASCE method using RZA-NLMF algorithm was proposed in [11]. The cost function of RZA-NLMF is given as

$$G_{RZA}(n) = \frac{1}{4}e^4(n) + \lambda_{RZA} \sum_{i=1}^{FIR} \log(1+\varepsilon|h_i(n)|), \quad (8)$$

where $\lambda_{RZA} > 0$ is a regularization parameter which trades off the estimation error and channel sparsity. The corresponding updated equation is derived as

$$\boldsymbol{h}(n+1) = \boldsymbol{h}(n) + \mu_N \frac{e(n)\boldsymbol{x}(n)}{\|\boldsymbol{x}(n)\|_2^2} - \rho \frac{\operatorname{sgn}(\boldsymbol{h}(n))}{1+\varepsilon|\boldsymbol{h}(n)|}, \quad (9)$$

where $\rho = \mu\lambda_{RZA}\varepsilon$ is a parameter which depends on step-size $\mu$, regularization parameter $\lambda_{RZA}$ and reweighted factor $\varepsilon$, respectively. In the second term of (10), if magnitude of $h_i(n)$, $i = 1,...,FIR$ is smaller than $1/\varepsilon$, these small coefficients will be replaced by zeros in high probability.

### 3.3. ASCE using RL1-NLMF (proposed)

The RL1-NLMF for sparse channel estimation has a better performance than the ZA-NLMF and RZA-NLMF which are usually employed in compressive sensing [11]. Because RL1-NLMF can exploit more channel sparsity information than ZA-NLMF as well as RZA-NLMF. Hence, the cost function of RL1-NLMF is devised as follows

$$G_{RL1}(n) = \frac{1}{4}e^4(n) + \lambda_{RL1} \|\boldsymbol{f}^T(n)\boldsymbol{h}(n)\|_1, \quad (10)$$

where $\lambda_{RL1}$ is the regularization parameter and $\boldsymbol{f}(n) = [f_1(n),\cdots,f_{FIR}(n)]^T$ is a reweighted vector where $f_i(n)$ is defined as

$$f_i(n) = \frac{1}{\delta + |h_i(n-1)|}, \quad i=1,2,\cdots,FIR \quad (11)$$

where $\delta$ should be some positive number, hence $f_i(n) > 0$. The updated equation can be derived by differentiating (11) with respect to the channel vector $\boldsymbol{h}(n)$. Then the resulting updated equation is:

$$\begin{aligned}\boldsymbol{h}(n+1) &= \boldsymbol{h}(n) + \mu_N \frac{e(n)\boldsymbol{x}(n)}{\|\boldsymbol{x}(n)\|_2^2} \\ &\quad - \mu\lambda_{RL1}\operatorname{sgn}(\boldsymbol{f}^T(n)\boldsymbol{h}(n))\boldsymbol{h}(n) \quad (12)\\ &= \boldsymbol{h}(n) + \mu_N \frac{e(n)\boldsymbol{x}(n)}{\|\boldsymbol{x}(n)\|_2^2} - \frac{\rho_{RL1}\operatorname{sgn}(\boldsymbol{h}(n))}{\delta + |\boldsymbol{h}(n-1)|},\end{aligned}$$

where $\rho_{RL1} = \mu\lambda_{RL1}$. In Eq. (12), since all signs of vector elements are one, i.e., $\operatorname{sgn}(\boldsymbol{f}^T(n)) = 1_{1\times N}$ with iteration times $n$, hence obviously can obtain result $\operatorname{sgn}(\boldsymbol{f}^T(n)\boldsymbol{h}(n)) = \operatorname{sgn}(\boldsymbol{h}(n))$. To evaluate the sparse penalty strength of ZA, RZA and RL1, corresponding three sparse penalty functions can be defined as below:

$$\zeta_{ZA} = \operatorname{sgn}(\boldsymbol{h}(n)), \quad (13)$$

$$\zeta_{RZA} = \frac{\operatorname{sgn}(\boldsymbol{h}(n))}{1+\varepsilon|\boldsymbol{h}(n)|}, \quad (14)$$

$$\zeta_{RL1} = \frac{\operatorname{sgn}(\boldsymbol{h}(n))}{\delta + |\boldsymbol{h}(n)|}, \quad (15)$$

where channel coefficients in $\boldsymbol{h}(n)$ are assumed in range $[-1,1]$. Considering above sparsity functions in Eqs. (13)~(15), their sparse penalty strength curves are depicted in Fig. 2. One can find that ZA utilizes uniform sparse penalty to all channel coefficients in the range of $[-1,1]$ and hence it is not efficient to exploit channel sparsity. Unlike the ZA (13), both RZA (14) and RL1 (15) make use of adaptively sparse penalty on different channel coefficients, i.e. stronger sparse penalty on zero/approximate zero coefficients and weaker sparse penalty on significant coefficients. Additionally one can also find that RL1 (15) utilizes stronger sparse penalty than RZA (14) as shown in Fig. 5. Hence RL1-LMF can exploit more sparse information than both ZA-LMF and RZA-LMF on adaptive sparse channel estimation. By virtue of Monte-Carlo (MC) based computer simulation, our proposed method will be verified in the following.

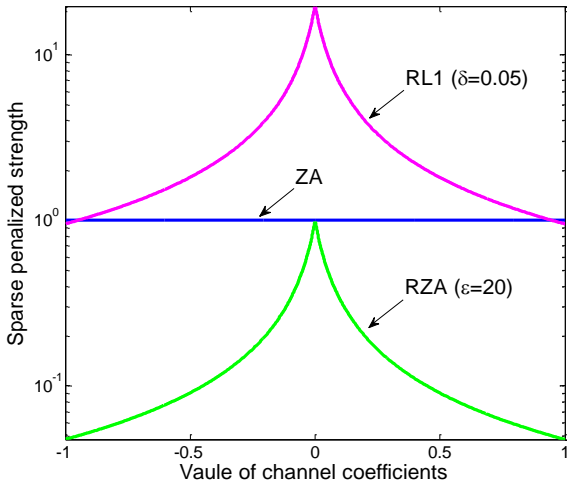

Fig. 5. Comparison of the three sparse penalty functions.

## 4. COMPUTER SIMULATIONS

In this section, the proposed ASCE method using RL1-NLMF algorithm is evaluated. For achieving average performance, $MC=100$ independent MC runs are adopted. The length of channel vector $h$ is set as $FIR=16$ and its number of dominant taps is set as $K \in \{1,4,6\}$. Each dominant channel tap follows random Gaussian distribution as $\mathcal{CN}(0,\sigma_h^2)$ and their positions are randomly allocated within the length of $h$ which is subject to $E\{\|h\|_2^2\}=1$. The received signal-to-noise ratio (SNR) is defined as $10\log_{10}(E_0/\sigma_n^2)$, where $E_0=1$ is the unit transmission power and $\sigma_n^2$ is noise variance. All of simulation parameters are listed in Tab. I. The estimation performance is evaluated by average MSD which is defined as

$$\text{Average MSD}\{h(n)\}=\frac{1}{MC}\sum_{m=1}^{MC}\|h_m(n)-h\|_2^2, \quad (16)$$

where $h$ denotes the actual channel vector and $h_m(n)$ stands for adaptive channel estimator at $m$-th MC run and $n$-th iteration. In the sequel, three simulation examples are shown to confirm the effectiveness of the proposed method.

Tab. I. Simulation parameters.

| Parameters | Values |
|---|---|
| Channel distribution of each dominant coefficient | Random Gaussian $CN(0,1)$ |
| Training sequence | Pseudorandom Binary training sequence |
| Channel length | $FIR=16$ |
| No. nonzero coefficients | $K \in \{1,4,6\}$ |
| Initial step-size | $\mu \in \{1.5, 2.0, 2.5\}$ |
| SNR | {8dB and 10dB} |
| Regularization parameters | $\lambda_{ZA}=5\times 10^{(3\sigma_n^2-5)/K}$ $\lambda_{RZA}=5\times 10^{(3\sigma_n^2-5)/K}$ $\lambda_{RL1}=5\times 10^{(3\sigma_n^2-8)/K}$ |
| Re-weighted factor of RZA-NLMF | $\varepsilon=20$ |
| Threshold of RL1-NLMF | $\delta=0.05$ |

**Example 1: MSD performance comparisons v.s. channel sparsity ($K$).**

To evaluate the performance of the proposed method, we compare it with other state-of-the-art methods, NLMF, ZA-NLMF, RZA-NLMF and RL1-NLMF, in Figs. 6~8. In the three figures, one can easily find that the proposed RL1-NLMF based ASCE method achieves lower MSD performance than previous three methods in different channel sparsity ($K$). For one thing, since ZA-NLMF based ASCE method may not exploit the channel sparsity effectively because MSD performance is very close to standard NLMF based ACE method. Hence, one can deduce that exploiting the sparsity could improve channel estimation accuracy. For another, RZA-NLMF can take more sparsity information than ZA-NLMF but the estimation performance can be further improved. Hence, the proposed method using RL1-NLMF can achieve better MSD performance than three previous methods due to the fact that it exploits channel sparsity efficiently. These simulation results are also coincidence with sparsity comparisons in Fig. 5. According to above discussion, one can find that the suitable channel sparsity is very useful on sparse channel estimation and other sparse system identification problems.

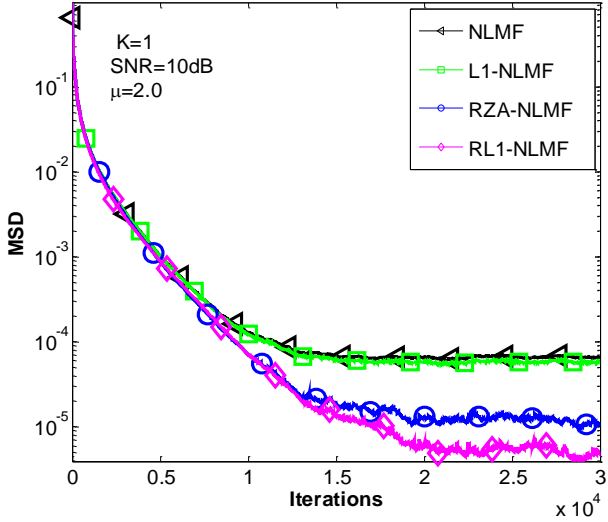

Fig. 6 MSD performance comparisons (*K*=1)

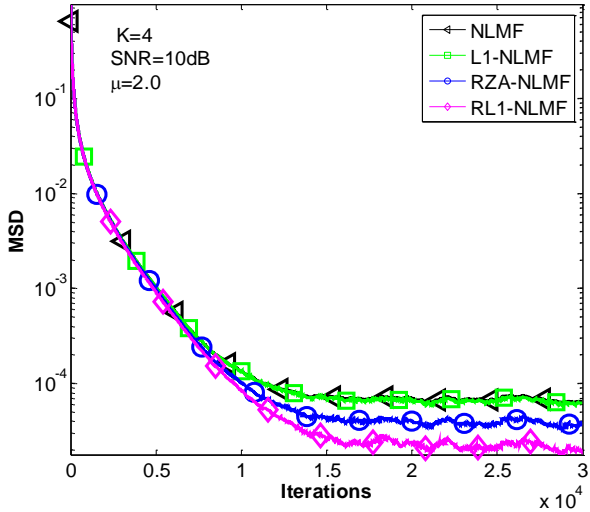

Fig. 7 MSD performance comparisons (*K*=4).

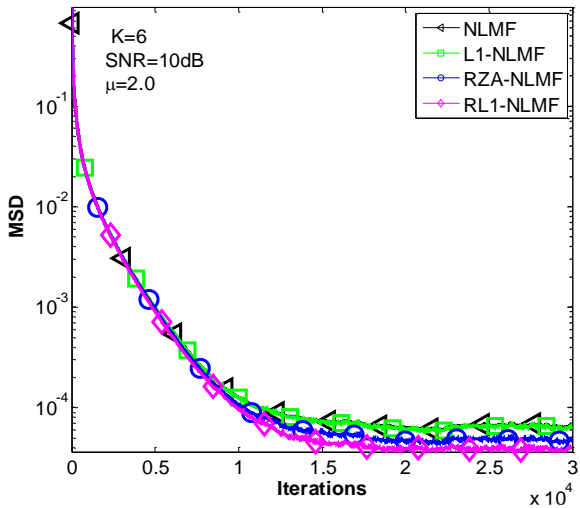

Fig. 8 MSD performance comparisons (*K*=6).

**Example 2: MSD performance comparisons v.s. initial step-sizes ($\mu$).**

Since the step-size is a critical parameter that decides the stability of adaptive filtering algorithm. In other words, it is necessary to evaluate the stability of the proposed method. Two initial step-sizes (i.e., $\mu = 2.5$ and 1.5) are adopted in adaptive filtering algorithm and performance curves are simulated in Figs. 9 and 10, respectively. According to the two figures, we can find that the proposed method can keep stable during gradient descend. In addition, one can find that initial step-size ($\mu$) may not change the convergence speed obviously because the step-size ($\mu_N$) depends on three factors: initial step-size ($\mu$), updating error ($e(n)$) as well as input training signal vector ($x(n)$). It is necessary to mention that suitable setting the initial step-size is still important. According to above discussion about the two figures, the stability of the proposed method by adopting different initial step-size can be confirmed.

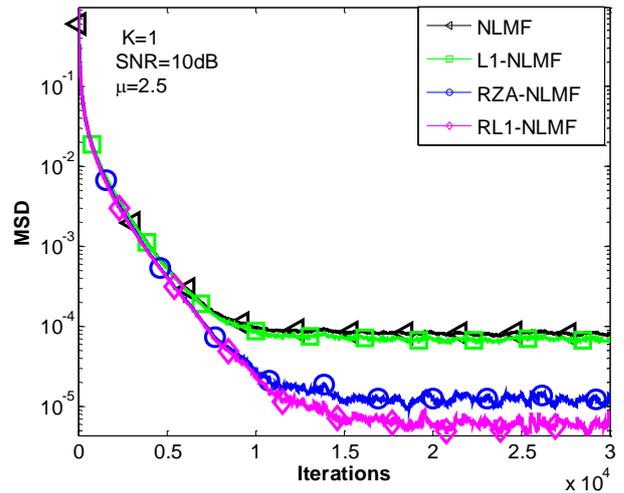

Fig. 9 MSD performance comparisons ($\mu = 2.5$).

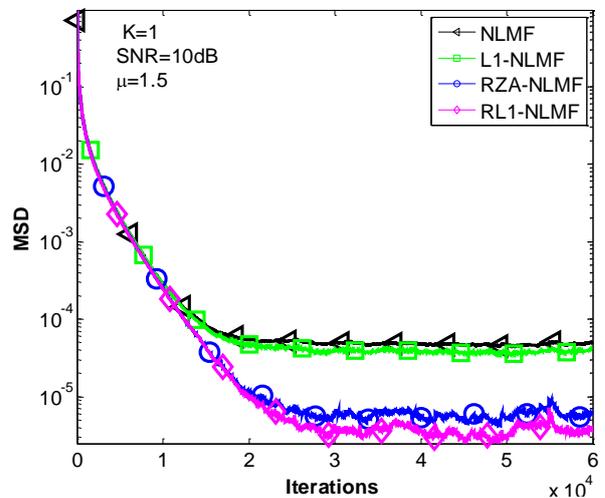

Fig. 10 MSD performance comparisons ($\mu = 1.5$).

**Example 3: MSD performance comparisons in the case of SNR=8dB.**

To further evaluate the performance of the proposed method, SNR=8dB is considered in Fig. 11. In this figure, one can find that the proposed method can achieve lower MSD performance than previous methods. In addition, the convergence speed of the NLMF-type methods is faster than case in SNR=10dB. According to this figure, we can deduce that convergence speed will be accelerated in low SNR environment due to the step-size will be enlarged in the case large updating error.

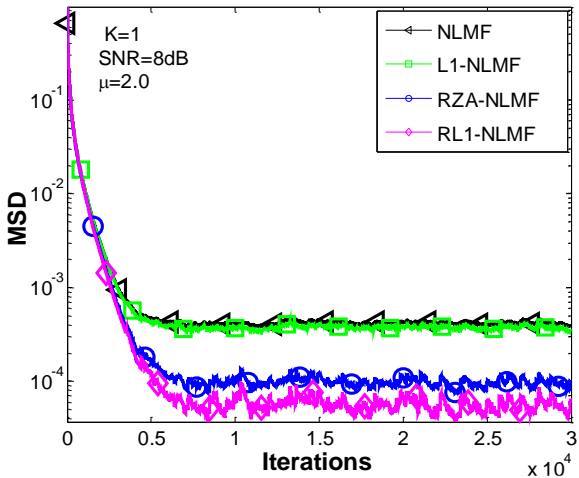

Fig. 11 MSD performance comparisons taps (SNR=8dB).

## 5. CONCLUSION AND FUTURE WORK

In this paper, an improved ASCE method using RL1-NLMF algorithm was developed. According to representative simulation results, our proposed method can achieve better MSD performance than ZA-NLMF and RZA-NLMF in different scenarios. Since this study is based on assumption of Gaussian noise model, it may unsuitable be applied in potential scenario in the presence of impulsive noise. The main reason is that existing sparse NLMF algorithms, e.g., RL1-NLMF, are unstable to impulsive noise. In future work, we are about to develop robust sparse NLMF algorithm to mitigate the impulsive noise in various sparse systems.

## REFERENCES

[1] L. Dai, Z. Wang, and Z. Yang, "Next-generation digital television terrestrial broadcasting systems: Key technologies and research trends," *IEEE Commun. Mag.*, vol. 50, no. 6, pp. 150–158, Sep. 2012.

[2] D. Raychaudhuri and N. B. Mandayam, "Frontiers of wireless and mobile communications," *Proc. IEEE*, vol. 100, no. 4, pp. 824–840, Apr. 2012.

[3] F. Adachi and E. Kudoh, "New direction of broadband wireless technology," *Wirel. Commun. Mob. Comput.*, vol. 7, no. 8, pp. 969–983, 2007.

[4] D. Tse, *Fundamentals of wireless communication*. Cambridge, U.K.: Cambridge University Press, 2005.

[5] E. Walach and B. Widrow, "The least mean fourth (LMF) adaptive algorithm and its family," *IEEE Trans. Inf. Theory*, vol. 30, no. 2, pp. 275–283, 1984.

[6] G. Gui and F. Adachi, "Adaptive sparse system identification using normalized least-mean fourth algorithm," *Int. J. Commun. Syst.*, vol. 28, no. 1, pp. 38-48, 2015.

[7] E. Eweda, "Global stabilization of the least mean fourth algorithm," *IEEE Trans. Signal Process.*, vol. 60, no. 3, pp. 1473–1477, Mar. 2012.

[8] L. Dai, Z. Wang, and Z. Yang, "Compressive sensing based time domain synchronous OFDM transmission for vehicular communications," *IEEE J. Sel. Areas Commun.*, vol. 31, no. 9, pp. 460–469, 2013.

[9] L. Dai, Z. Wang, and Z. Yang, "Spectrally efficient time-frequency training OFDM for mobile large-scale MIMO systems," *IEEE J. Sel. Areas Commun.*, vol. 31, no. 2, pp. 251–263, 2013.

[10] Z. Gao, L. Dai, Z. Lu, C. Yuen, and Z. Wang, "Super-resolution sparse MIMO-OFDM channel estimation based on spatial and temporal correlations," *IEEE Commun. Lett.*, vol. 18, no. 7, pp. 1266–1269, 2014.

[11] G. Gui, L. Xu, and F. Adachi, "RZA-NLMF algorithm-based adaptive sparse sensing for realizing compressive sensing," *EURASIP J. Adv. Signal Process.*, vol. 2014, p. 125, 2014.

[12] E. J. Candes, M. B. Wakin, and S. P. Boyd, "Enhancing Sparsity by Reweighted l1 Minimization," *J. Fourier Anal. Appl.*, vol. 14, no. 5–6, pp. 877–905, 2008.

[13] G. Gui and F. Adachi, "Sparse least mean forth filter with zero-attracting," *Submitt. IEICE Electron. Express*, pp. 1–6, 2013.

[14] D. L. Donoho and Y. Tsaig, "Fast solution of L1-norm minimization problems when the solution may be sparse," *IEEE Trans. Inf. Theory*, vol. 54, no. 11, pp. 4789–4812, 2008.

[15] G. Gui, W. Peng, and F. Adachi, "Improved adaptive sparse channel estimation based on the least mean square algorithm," in *IEEE Wireless Communications and Networking Conference (WCNC), Shanghai, China, 7-10 April*, 2013, pp. 3130–3134.